\documentclass[twocolumn,prd,amsmath,amssymb]{revtex4}

\usepackage{graphicx}
\usepackage{dcolumn}
\usepackage{bm}

\begin{document}

\title{Nucleon spin structure studies at Jefferson Lab}
\author{A.~Deur}

\affiliation{
\baselineskip 2 pt
\hspace{1mm}
\centerline{{Thomas Jefferson National Accelerator Facility}}
}

\date{\today}

\begin{abstract}
We report on the low and moderate $Q^2$ nucleon spin structure measurements done at Jefferson Lab, examining specifically the
inclusive program. We discuss what the data teach us about quark confinement and the
emergence of the effective hadronic degrees of freedom from the fundamental partonic ones.
We show how this experimental program has reached its goal by providing a precise mapping at  low, intermediate and
moderately high $Q^2$ which has followed in many advances, e.g., with Chiral Perturbation Theory.
Another example of a recent advance imputable to the JLab spin data is  the improved understanding of 
$\alpha_s$ at low energy, which allowed Light--Front Holographic QCD, an approximation to non-perturbative QCD,
to  derive the hadron spectrum from $\Lambda_s$. 
\end{abstract}

\maketitle

\section{Introduction}
\vspace{-0.3cm}
There are many reasons to study the spin of the nucleon. An ingenuous reason is that it is interesting in itself: we are curious to know
how the spin  of the nucleon emerges from its constituents, i.e., how the spins and orbital angular momenta
of the quarks and gluons combine together into the spin  1/2  of the nucleon. We have also more practical reasons: spin degrees
of freedom provide additional tests to theories. Hence, studying spin observables provides a more complete study of the fundamental
force ruling the nucleon structure, the Strong Force, of which the accepted gauge theory is Quantum Chromodynamics (QCD). 
We can also learn about important emergent properties
of QCD, such as the mechanism of confinement, how the effective hadronic degrees of freedom stem from the fundamental 
partonic ones, or how well non-perturbative  methods, e.g., Chiral Perturbation Theory ($\chi$PT), Lattice Gauge Theory
or the gravity-gauge (AdS/CFT) correspondence, describe the Strong Force in its low energy domain. Finally, a precise
knowledge of polarized parton distributions is necessary for investigations of physics beyond the standard model or high precision
atomic physics. We have thus many good reasons to be interested in the nucleon spin. 

In this document, we will focus on what spin structure studies teach us about confinement, the emergence of the hadronic degrees of 
freedom  and tests of non-perturbative approaches to QCD. 
Due to space limits,  we will discuss results based on inclusive
lepton scattering, which are only a fraction of the polarized observables necessary for a full description of the nucleon spin structure.
In fact, we will focus on the first moment of the spin structure function $g_1$ at low to moderate $Q^2$. 
These results were obtained with the 6 GeV JLab accelerator. The future spin program with the new 12 GeV JLab beam
is discussed in V. Burkert's contribution to these proceedings.   
\section{Inclusive polarized  lepton scattering}
Inclusive polarized  lepton scattering data, obtained by experiments in which the only particle of the reaction 
detected is the scattered lepton, is parametrized (neglecting the weak interaction) by four structure functions: 
two unpolarized ones: $F_1(x,Q^2)$ and $F_2(x,Q^2)$, and two polarized ones: $g_1(x,Q^2)$ and $g_2(x,Q^2)$ 
($x$ is the Bjorken scaling variable and $Q^2=-q^2$, with $q \equiv (\nu,\overrightarrow{q})$ the 4-momentum exchanged between the beam 
particle and the target particle). In the Bjorken limit, $Q^2\rightarrow\infty$ while $Q^2 / \nu$ stays finite, $F_1(x)$ 
and $F_2(x)$ are constructed from the quark densities 
$u(x)$($\overline u(x)$), $d(x)$($\overline d(x)$), $s(x)$($\overline s(x)$)... which provide the amount of up, 
down, strange... quarks (antiquarks) carrying a nucleon momentum fraction $x$, respectively. $F_1(x)$ and 
$F_2(x)$ are simple sums of these densities: $F_1=q_u^2[u + \overline u]+q_d^2[d + \overline d]+...$ and 
$F_2=2xF_1$, with  $q_f$ the electric charge of the quark of flavor $f$. Likewise, $g_1(x)$ 
is constructed from the quark polarizations $\Delta u(x)$... which are, for a given $x$, the difference between the amount of 
quarks with their spins parallel to the nucleon spin and those with their spins antiparallel:
$g_1=q_u^2[\Delta u + \Delta \overline u]+q_d^2[\Delta d + \Delta \overline d]+...$
In the Bjorken limit the other spin structure function has no role: $g_2 = 0$.

This relatively simple picture is only valid at $Q^2 \rightarrow \infty$ where the QCD coupling $\alpha_s$ should vanish,
i.e. in the asymptotic case where quarks are free. At finite (but still large) $Q^2$, quarks start to interact, which leads
to gluon corrections. Furthermore, the mass of the target  $M$ and effects of transverse momentum $P_{\perp}$ or 
transverse spin are less suppressed: $M^2/Q^2 \neq 0$, $P_{\perp}^2/Q^2 \neq 0$ and thus, they start to play a role. 
This breaks down the above relations: $F_2 \neq 2xF_1$ and $g_2 \neq 0$. 

At smaller $Q^2$, ($\simeq$ GeV$^2$), $\alpha_s$ becomes large and the significantly stronger quark interaction 
induces correlations, known as Higher Twists (HT) effects, which invalidate the simple description summarized above: 
$F_1 \neq q_u^2[u + \overline u]+q_d^2[d + \overline d]+...$ and 
$g_1 \neq q_u^2[\Delta u + \Delta \overline u]+q_d^2[\Delta d + \Delta \overline d]+...$. 
Some of the HT effects that contribute to inclusive polarized scattering have been interpreted
recently as the transverse confining force acting on the quarks~\cite{Burkardt:2008ps}. HT have been
measured in a number of Jefferson Lab (JLab) experiments in Halls A~\cite{A1n-g2n-Posik}, B~\cite{EG1} and C~\cite{RSS}.
In particular, the HT were extracted in the high-$x$ domain in~\cite{LSS}. The SANE experiment
in Hall C will further add to these data (see W. Armstrong's contribution to these proceedings).

An example of HT extraction specially relevant to this document is the determination of the twist-4 element $f_2^{p-n}$~\cite{EG1}.
It is extracted from Bjorken sum measurements (see next section) at JLab. Its absolute value at $Q^2=1$ GeV$^2$, 
$f_2^{p-n} = -0.064(35)$, is large (the scale is set by the leading twist value of the Bjorken sum, $\Gamma_1^{p-n, LT} = 0.137(13))$).
Such large HT agrees with the  intuition that non-perturbative effects should be large at moderate $Q^2$. However, the $1/Q^4$ HT correction 
$\mu_6$ is small and $\mu_8$ is of similar magnitude as $f_2$ but of opposite sign. Thus, overall the sum of HT is small near  
$Q^2=1$ GeV$^2$. This explains why HT have been so elusive, as well as the onset of hadron-parton 
duality~\cite{Bloom:1970xb}. If HT are overall small at these $Q^2$, what are their actual effects and what is their practical
connection to confinement? To answer these questions, we first need to discuss the subject of ``sum rules".

\section{Sum rules}

\begin{figure*}[ht!]
\centering
 \includegraphics[width=0.95\textwidth]{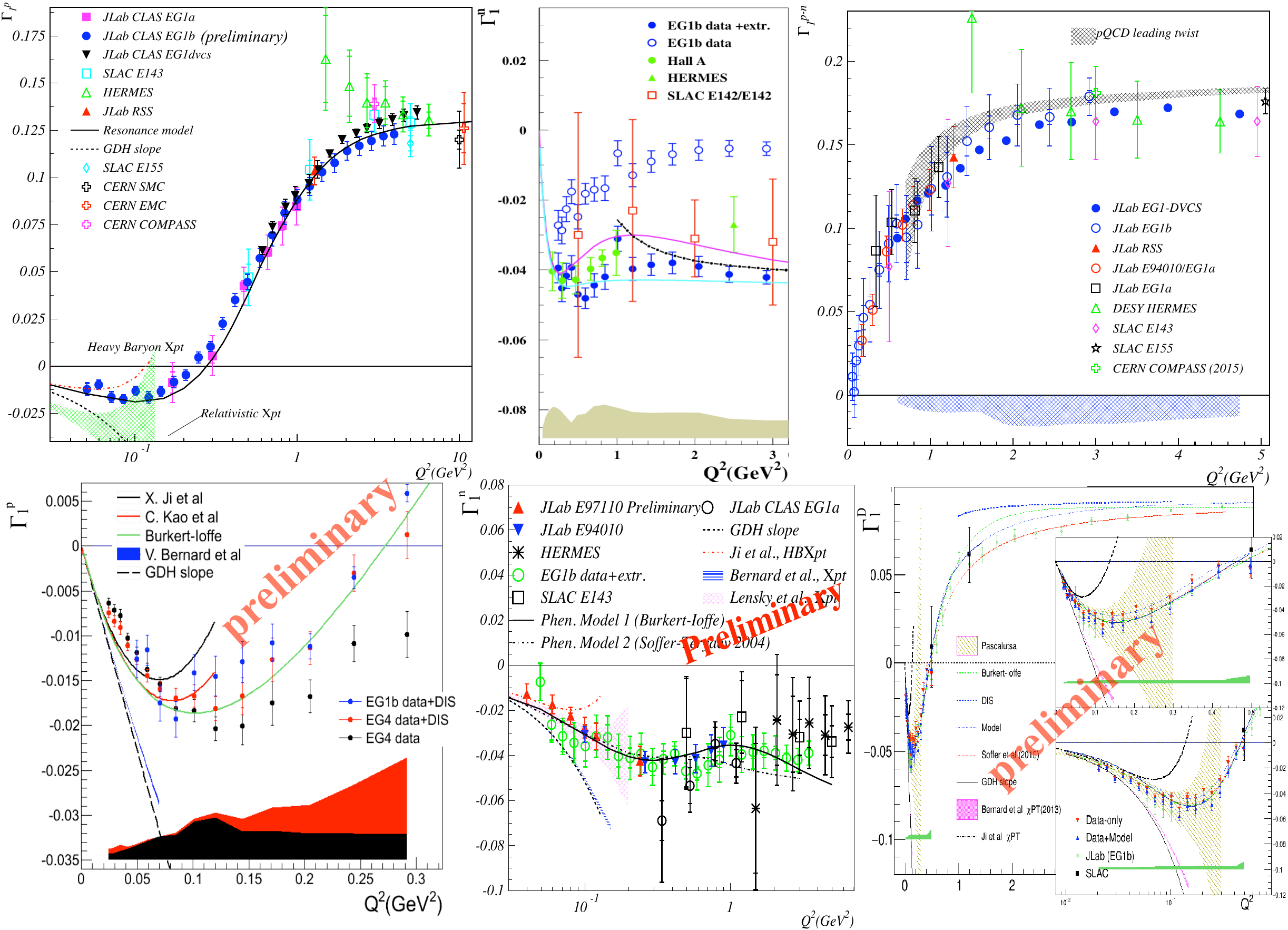}
\caption{Top panels: Measurements of  $\Gamma_1^p$ (left), $\Gamma_1^n$ from D (center) and $\Gamma_1^{p-n}$ (right). Bottom panels:  
preliminary low $Q^2$ results from JLab's experiments EG4 (p: left; n from D: right) and E97110 (center, n from $^3$He).
\label{Fig1} }
\end{figure*}

A Sum Rule is a rule (i.e an equality) that relates a sum (i.e., an integral such as a moment of a structure function) to a 
quantity characterizing the studied particle. A relevant sum rule to this document is the Gerasimov-Drell-Hearn (GDH) sum rule \cite{Gerasimov:1965et}
that relates the photo-absorption cross-sections $\sigma^{3/2}$ and $\sigma^{1/2}$ (3/2 (1/2) indicates that the photon helicity 
is aligned (anti-aligned) with the target polarization) to the anomalous magnetic moment $\kappa_t$ of the target:
\begin{equation} 
\int_{\nu_{thr}}^{\infty} \big(\sigma^{3/2}(\nu)-\sigma^{1/2}(\nu) \big) \frac{d\nu}{\nu} = \frac{2\alpha \pi^2 \kappa_t^2}{M^2}, \label{GDH}
\end{equation} 
with $\nu$ the photon energy, $\nu_{thr}$ the photo-production threshold, $\alpha$ the QED coupling constant and $M$ the target mass.
Originally derived for photo-production ($Q^2=0$), the GDH sum rule has been latter generalized to $Q^2>0$~\cite{Ji:1999mr}:
\begin{equation} 
\Gamma_1 \equiv \int_{0}^{1} g_1(x,Q^2)dx = \frac{Q^2}{8}S_1, \label{GDH general}
\end{equation} 
where $S_1$ is a double DVCS spin--dependent amplitude.
The famous Bjorken sum rule for spin-dependent deep inelastic scattering (DIS)~\cite{Bjorken:1966jh}  then appears as the isovector part 
of the generalized GDH sum rule in the large $Q^2$ limit:
\begin{equation} 
\int_{0}^{1} (g_1^p-g_1^n)dx = \frac{1}{6}\bigg(1-\frac{\alpha_s}{\pi}-3.58\big(\frac{\alpha_s}{\pi}\big)^2... \bigg) + 
\sum_{i=2}\frac{\mu_{2i}}{Q^{2i}}, \label{bjorken SR}
\end{equation} 
where the series coefficients and $\alpha_s$ are expressed in the $\overline {MS}$ renormalization scheme (RS), and where
the $\mu_{2i}$ are non-perturbative HT corrections --discussed in the preceding section-- that become important only at low $Q^2$.
Data on $\Gamma_1^p$, $\Gamma_1^n$ and $\Gamma_1^{p-n}$, including  preliminary results from experiments 
EG1, EG4 and E97110 are shown in Fig. \ref{Fig1}. 
The high precision of the mapping at  intermediate and moderately--high $Q^2$ done for the proton and the neutron 
at JLab can be seen on Fig. \ref{Fig1} (top plots). The  $\Gamma_1(Q^2)$ moments display a strong $Q^2$--variation
while transiting from high-to-low $Q^2$. The mapping is being finalized with the analysis of more recent
experiments covering the $\chi PT$ domain at very low $Q^2$ (preliminary results shown on the bottom plots). 
Again, the data have high precision and should test well the $\chi PT$ predictions. The preliminary results on low-$Q^2$ measurement of the 
generalized GDH sum rule on the neutron($^3$He) are discussed in C. Peng's contribution to these proceedings. 
We see that the recent $\chi PT$ results agree with the preliminary data.
We can also notice the similar negative slopes
for $\Gamma_1^p$ and $\Gamma_1^n$ at very low $Q^2$. This is explained by the similar absolute values of the anomalous magnetic moments
of the proton and neutron (assuming the GDH sum rule is not violated, or at least not strongly). Consequently, these slopes
mostly cancel in the Bjorken sum $\Gamma_1^{p-n}$, which results for this observable in a restoration of the approximate
conformal behavior (no $Q^2$-dependence) of QCD already seen at large $Q^2$. This observation is important in the context
of using the AdS/CFT (Anti-de Sitter/\emph{Conformal} Field Theory) approach to non-perturbative QCD, as we will discuss latter.

Other spin-dependent sum rules exist, such as the Burkhardt-Cottingham sum rule on $g_2$ \cite{Burkhardt:1970ti} or polarizabilities sum rules
that involve higher moments of $g_1$ and $g_2$, see C. Peng, J. Zhang and K. Slifer contributions to these proceedings.

The high $Q^2$ data have historically provided a check of QCD in its spin sector, as well 
as the measurement of the quark spin contribution to the nucleon spin and --\emph{via} evolution equations-- a first assessment of the gluon 
spin contribution $\Delta$G (see e.g.~\cite{LSS}). The moderate $Q^2$ data provide --as discussed in the previous section-- a handle 
on HT. The new low-$Q^2$ data test recent $\chi$PT calculations~\cite{chipt},
see C. Peng and K. Slifer contributions to these proceedings. We will discuss now
what we learn in practice from these measurements.

\section{Insights in the non-perturbative nucleon structure}

Exploring the connection between hadronic and partonic degrees of freedom was the main goal of the JLab experiments that covered the 
moderate $Q^2$--values that bridge the hadronic and partonic domains. That is, their goal was to provide data to help developing/improving 
non-perturbative theoretical frameworks, and to help explain how these various theoretical frameworks 
are connected. We will show here how this goal was achieved.

The precise mapping of low and moderate $Q^2$ regions motivated:
\begin{enumerate} 
\item improvement at low $Q^2$ on $\chi$PT calculations~\cite{chipt}, in particular to 
address the $\delta_{LT}$ puzzle, see C. Peng, J. Zhang  and K. Slifer contributions to these proceedings;
\item improvement at higher $Q^2$ of pQCD
techniques, such as the Analytic or Massive Perturbation Theories~\cite{impr pQCD}. These approaches essentially fold partially the 
HT contributions into the definition of $\alpha_s$~\cite{Deur:2016tte}  in order to extend the domain of applicability of pQCD series to lower $Q^2$; 
\item the calculation of $\alpha_s$ using an AdS/CFT duality approach~\cite{Brodsky:2010ur} . 
\end{enumerate} 
This is this last item that we will discuss to illustrate how the JLab spin data have pushed fundamental advances in our understanding of
non-perturbative QCD.

\section{The Light--Front Holographic QCD approximation}

The Light--Front Holographic QCD (LFHQCD) approximation to QCD~\cite{LFHQCD review} is based on light-front quantization, which allows a 
rigorous and exact formulation of non-perturbative QCD~\cite{Brodsky:1997de}. In particular, light-front  quantization yields a 
relativistic Schr\"{o}dinger-like equation describing hadrons as quark bound-states. 
In principle, all elements of this equation can be determined from the QCD Lagrangian; but, in
practice, it is only in (1+1) dimensions that we know how to compute the effective confining potential entering the equation~\cite{Hornbostel:1988fb}.
Due to the overwhelming complexity, in (3+1) dimensions, the potential must be determined from other means than first-principle light-front calculations. One such method
is  the correspondence between gravity in anti-de Sitter (AdS) space and QCD on the light-front~\cite{AdS/QCD}. 
This correspondence originates from the fact that the group of isometries of a 5-dimensional AdS space is the conformal invariance of the dual field theory in Minkowsky space, and thus encodes the scale invariance of the classical QCD Lagrangian. The calculations are tractable
if short-distance quantum fluctuations are neglected and if the quark masses are set to zero (chiral limit). The 5-dimensional AdS calculations are 
projected on the 4-dimensional boundary of the AdS space which is identified with the physical 
Minkowski spacetime (hence the ``holographic" denomination). 
Thus, Light-Front Holography provides a semiclassical approximation to QCD in its long distance regime which incorporates fundamental aspects of QCD.

Enforcing the conformal symmetry of QCD, i.e that the Lagrangian has no energy scale in the chiral limit, 
without explicitly breaking the symmetry, leads to a unique choice of potential~\cite{unique HO}. It has the 
form of a harmonic oscillator on the light front. The uniqueness of this form is confirmed by the facts that 
1) only this form yields a zero-mass pion in the chiral limit~\cite{Dosch:2015nwa}; 
2) it explains the remarkable mass symmetry between mesons and baryons~\cite{deTeramond:2014asa};
3) it is equivalent to the well-established linear potential for static quarks  in the usual instant-form front~\cite{Trawinski:2014msa}.
 
Hence, fundamental aspects of nonperturbative QCD  are incorporated in LFHQCD which is based on the embedding of light-front relativistic bound state equations in AdS space. This semiclassical approximation to QCD in its large distance domain is completely determined by the constraints imposed by superconformal algebra, a symmetry which can originate from the dynamics of color SU(3)~\cite{deTeramond:2016htp}. The only free parameter is the scale $\kappa$. In fact, chiral QCD is independent of  conventional units of mass such as MeV. 
In other words, a
theory or model of the Strong Force can only predict dimensionless ratios such as the proton 
to $\rho$--meson mass ratio $M_p / M_{\rho}$, or  $M_p / \Lambda_s$.
For LFHQCD this single parameter is denoted $\kappa$. For standard QCD, it is $\Lambda_s$. The relation between the two is
known analytically and numerically~\cite{Deur:2014qfa} and was obtained \emph{via} the QCD coupling $\alpha_s$, which we discuss next.

\section{The QCD coupling $\alpha_s$}

A quantity of central interest for this article that can be determined by LFHQCD is the strong 
coupling $\alpha_s$~\cite{Deur:2016tte}. A particularly useful way to define
it in the long distance regime is \emph{via} the concept of effective charge~\cite{Grunberg}, which is defined from the perturbative series of an osbservable, truncated to first order in $\alpha_s$. For example, with such coupling, Eq.~(\ref{bjorken SR}) becomes:
\begin{equation} 
\int_{0}^{1}( g_1^p-g_1^n)dx \equiv \frac{1}{6}\big(1-\frac{\alpha_{g_1}}{\pi}\big), \label{alpha_g1}
\end{equation} 
where the subscript $g_1$ reminds us of this particular choice of definition. 
This choice can be viewed as equivalent to a RS choice~\cite{Deur:2016cxb}.
One sees that with this definition, both the short distance pQCD effects (the higher order terms in the leading twists 
pQCD series) and long distance confinement 
effects (the HT terms) are now folded into the definition of $\alpha_s$.  This is in analogy with the original coupling constant 
becoming a running (effective) coupling when short distance quantum effects are folded into its definition~\cite{Deur:2016tte}.
The folding of the long distance confinement effects into the coupling definition regularizes it, removing the unphysical Landau pole at 
$Q^2=\Lambda_s^2$~\cite{Deur:2016tte}. We already mentioned how the folding of HT lead to optimized perturbative series.
The definition of $\alpha_s$ as an effective charge can be viewed as a generalization of this procedure:  the role of HT --or more generally 
the long distance confining effects-- is to regularize $\alpha_s$. 

\begin{figure}[ht!]
\centering
 \includegraphics[width=0.4\textwidth]{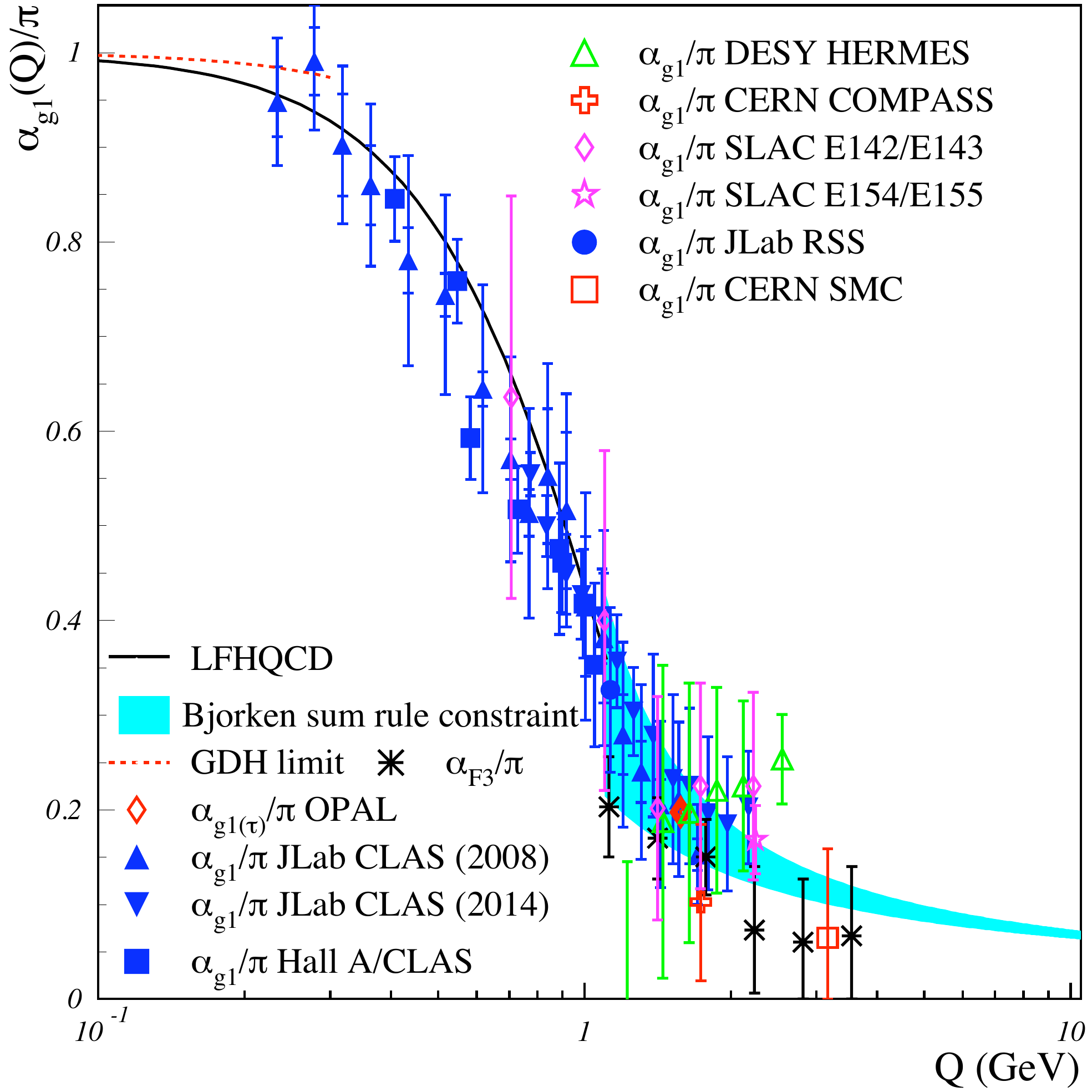}
\caption{The effective charge $\alpha_{g_1}/\pi$ from experimental data (symbols). The data are supplemented at low $Q^2$
by the GDH sum rule (dashed line) and at high $Q^2$ by the Bjorken sum rule (band). The LFHQCD prediction is given by the 
continuous line.
\label{fig:alpha_g1}
\protect \\
}
\end{figure}

The Bjorken sum measurement can be used to obtain $\alpha_{g_1}$~\cite{Deur:2005cf}. The most 
important data are from JLab at moderate and low $Q^2$. At lower and higher $Q^2$, the GDH and Bjorken sum rules, respectively, 
can  supplement the data. The experimental data yield the coupling shown in Fig.~\ref{fig:alpha_g1}. 
%
\begin{figure*}[ht!]
\centering
\centerline{\includegraphics[width=.35\textwidth]{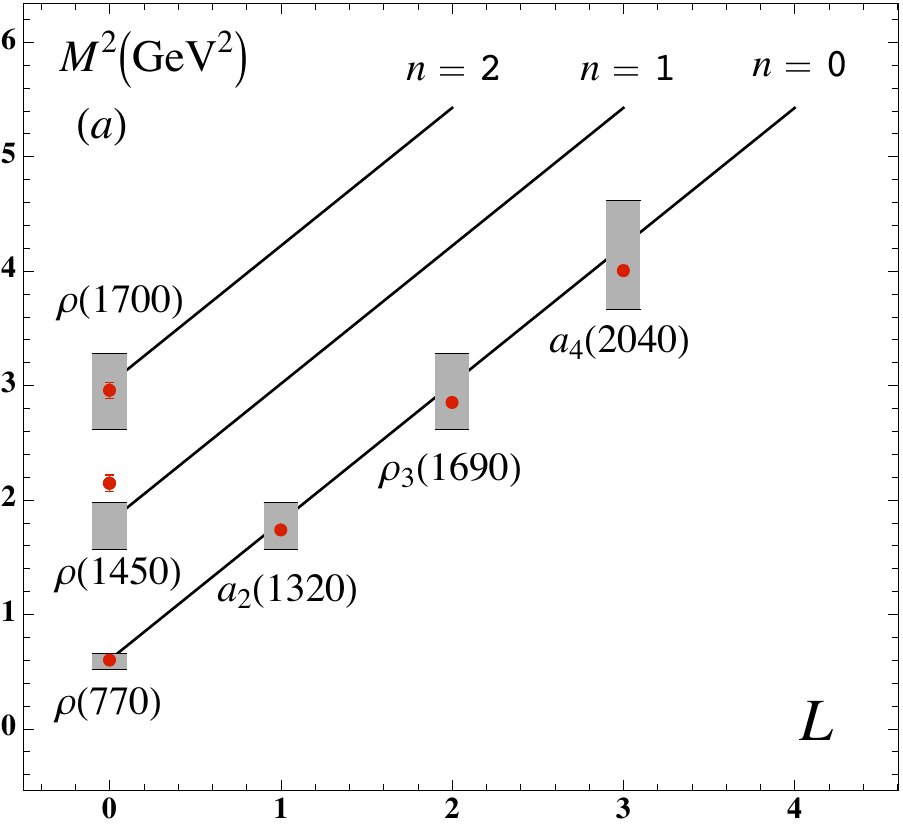} 
\includegraphics[width=.35\textwidth]{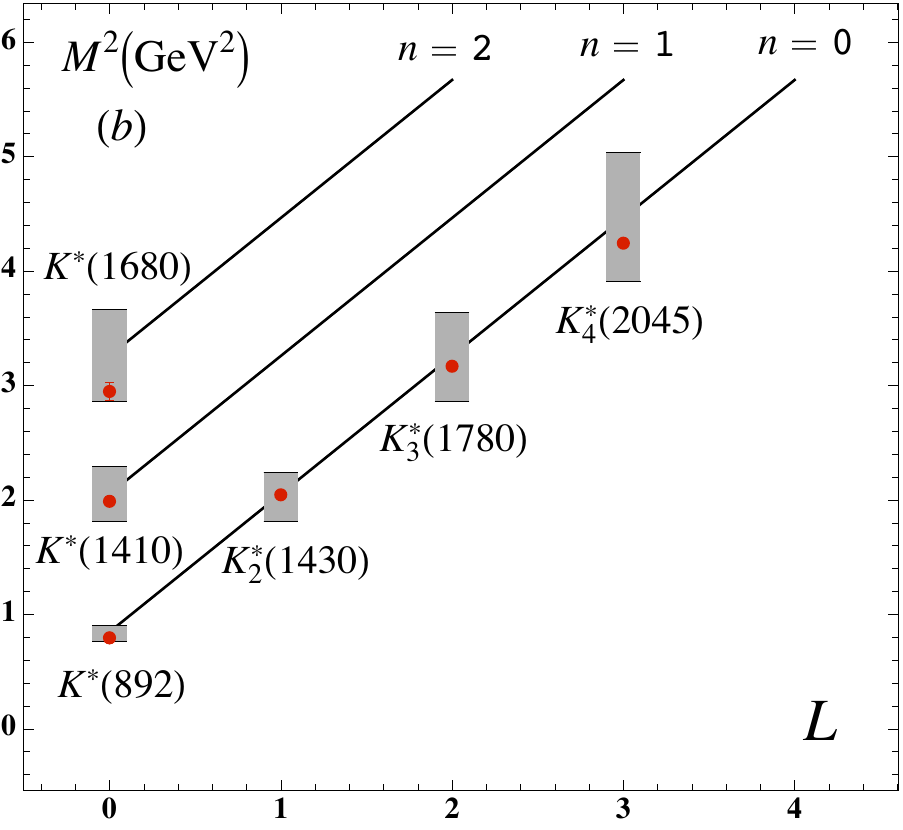}}
\caption{\label{Fig:masses} The predicted mass spectrum from LFHQCD for unflavored (a) and strange light vector mesons (b). 
The gray bands represent the uncertainty. The points are the experimental values.}
\end{figure*}
An important observation provided by the measurement of $\alpha_{g_1}$ at low $Q^2$ is that it freezes, i.e., QCD's 
approximate conformal behavior is restored. (QCD is also conformal at large $Q^2$, a phenomenon called Bjorken Scaling).
This allows us to apply LFHQCD at low $Q^2$. It predicts that ~\cite{Brodsky:2010ur}:
\begin{equation} 
\alpha_s(Q^2) = \pi e^{-\frac{Q^2}{4\kappa^2}}.\label{alpha_s from LFHQCD}
\end{equation} 
This prediction agrees well with the data, see Fig.~\ref{fig:alpha_g1}. It is important to remember that there are no free parameters
in Eq.~(\ref{alpha_s from LFHQCD}), $\kappa$ being determined by hadron masses (in  Fig.~\ref{fig:alpha_g1},
$\kappa=M_{\rho}/\sqrt{2}$~\cite{LFHQCD review}). The LFHQCD prediction is drawn in Fig.~\ref{fig:alpha_g1} up to 
$Q^2 \simeq 1$ GeV$^2$, its expected domain of validity. At higher $Q^2$, short distance quantum effects, which are not included in the usual semiclassical approximation in LFHQCD become important and void the prediction. There, however, pQCD is available to calculate $\alpha_{g_1}$.
In fact, the respective domains of applicability of LFHQCD and pQCD appear to overlap around $Q^2 \simeq 1$ GeV$^2$~\cite{Deur:2016cxb}.
This allows us to match the $\alpha_{g_1}$ computation from pQCD to that from LFHQCD, thereby relating hadronic masses 
to the fundamental QCD parameter $\Lambda_s$~\cite{Deur:2014qfa}. For example, the leading order relation between $\Lambda_{\overline{MS}}$
and $M_\rho$ is:
\begin{equation}  \label{eq: Lambda LO analytical relation}
\Lambda_{\overline{MS}}=M_\rho e^{-a}/\sqrt{a},
\end{equation} 
where $a=4\big(\sqrt{ln(2)^{2}+1+\beta_{0}/4}-ln(2)\big)/\beta_{0}$, ($\beta_{0}=11-2n_f/3$ is the first coefficient of the QCD's $\beta$-series). 
Numerically, $a\simeq 0.55$ for $n_{f} = 3$ quark flavors. At N$^3$LO the 
numerical relation is $\Lambda_{\overline{MS}}=0.440 M_\rho$. The $\rho$ meson is the ground state solution of the LFHQCD
Schr\"{o}dinger equation, i.e., with radial excitation $n=0$ and internal orbital angular momentum $L=0$. Higher mass states are 
solutions with $n \neq 0$ and $L \neq 0$ and are shown on Fig.~\ref{Fig:masses}, together with a similar prediction for strange mesons.
Baryon masses can be obtained similarly or  thanks to the mass symmetry between mesons and baryons~\cite{deTeramond:2014asa}.
Conversely, the known value of $\kappa$ can be used  with the same matching procedure to predict $\Lambda_s$. This leads to 
$\Lambda_{\overline{MS}}^{n_f=3}=0.339(19)$ GeV, which agrees well with the world average value of 0.339(17) GeV~\cite{Olive:2016xmw}.

\section{Summary and conclusion}
We have reported on the low and moderate $Q^2$ (few GeV$^2$) nucleon spin structure measurements done at JLab, focusing on the
inclusive program on the experimental side. On the phenomenology side, we focused on what this teaches us about quark confinement and the
emergence of effective degrees of freedom (hadrons) from the fundamental ones (quark and gluons). 
The JLab data cover an extensive kinematic range with high precision thanks to the high luminosity of the experimental equipment.
They complement the data from the CERN, SLAC and DESY high energy facilities. 
Most of the data at moderate $Q^2$ are now available. Remaining data from the CLAS EG1b and Hall C SANE experiments 
will be available soon. The analysis of the lower $Q^2$ data is being finalized for the neutron, both from polarized deuteron and $^3$He 
targets, and should be available early 2017.  More work remains for the low $Q^2$ proton data. Finally, some very low $Q^2$ data 
($\approx 0.01<Q^2<0.04$ GeV$^2$) from 
n~($^3$He) should become available in the upcoming years.
  
We argued that the JLab spin sum rule program has reached its goal: it has provided a precise mapping at  low, intermediate and
moderately high $Q^2$ and triggered advances on the theoretical front. In particular, there is  good progress in the description of the 
Strong Force over the full $Q^2$-range, with improvements in the Chiral Perturbation Theory calculations (low $Q^2$) and 
pQCD series (high $Q^2$).
A goal for these data was to help understanding how hadronic degrees of freedom connect to partonic ones. 
Better measurements and understanding of Higher Twists, as well as new approaches to the confinement problem, e.g., from Light--Front 
Holographic QCD, represent advances on this front. 
As examples of recent developments directly connected to the JLab spin data, we showed the extraction of the QCD coupling
$\alpha_s$ and its determination from Light--Front Holographic QCD, whose applicability domain overlaps
with that of pQCD near $Q \simeq$ 1 GeV. This overlap allows the analytical determination of the hadron spectrum with $\Lambda_s$ as the only input. 
One parameter e.g., $\Lambda_s$ allows one to express quantities in terms of conventional GeV units. 
Since $\Lambda_s$ is well determined, the mass spectrum prediction  
from Light--Front Holographic QCD has no free parameters. Obtaining such result has been 
the long-thought goal of the Strong Force studies. 
Arguably Light-front holographic QCD is not QCD, but it is a semiclassical approximation which incorporates basics aspects of nonperturbative QCD dynamics which are not apparent from the QCD Lagrangian, such as the emergence of a mass scale and confinement, the existence of a zero mass particle in the chiral limit and universal Regge trajectories~\cite{deTeramond:2016htp}. Hence the derivation of the strong coupling in the infrared domain and the hadron spectrum represents an important advance 
toward reaching this long-thought goal, and the Jefferson Lab spin data has played a major role in it. 

\begin{acknowledgements}
We thank  S.~J.~Brodsky, E.~Chudakov, V.~Sulkosky, G.~F.~de T\'{e}ramond and N.~Ton for their useful comments on the manuscript.
This material is based upon work supported by the U.S. Department of Energy, Office of Science, Office of Nuclear Physics under contract DE--AC05--06OR23177. 
\end{acknowledgements}

\end{document}